\documentclass{article}
\usepackage[preprint]{spconfa4}
\usepackage{amsmath,graphicx}
\usepackage{amssymb}
\usepackage{tikz}
\usepackage{stackengine}        
\usepackage{cite}               

\usepackage{lipsum}         

\usepackage[subtle]{savetrees}
\usepackage[hidelinks]{hyperref}

\usepackage{balance} 

\newcommand{\y}{\mathbf{y}}
\newcommand{\x}{\mathbf{x}}
\newcommand{\n}{\mathbf{n}}
\newcommand{\h}{\mathbf{h}}
\newcommand{\hest}{\tilde{\h}}
\newcommand{\hestSC}{\tilde{\h}^{\rm SC}}
\newcommand{\hestMAX}{\tilde{\h}^{\rm mSNR}}
\newcommand{\Hmat}{\mathbf{H}}
\newcommand{\Hest}{\tilde{\Hmat}}
\newcommand{\Emat}{\mathbf{E}}
\newcommand{\esub}{\mathrm{e}}
\newcommand{\Me}{M_\esub}

\newcommand{\asub}{\mathrm{a}}
\newcommand{\Ma}{M_\asub}

\newcommand{\ones}{\mathbf{1}}
\newcommand{\zeros}[2]{\mathbf{0}_{#1\times#2}}
\newcommand{\avec}{\boldsymbol{\alpha}}
\newcommand{\Rn}{\mathbf{R}_\mathrm{n}}
\newcommand{\Rnest}{\hat{\mathbf{R}}_\mathrm{n}}

\newcommand{\rn}[1]{R_{\mathrm{n},#1}}
\newcommand{\Rninv}{\mathbf{R}_\mathrm{n}^{-1}}

\newcommand{\Ry}{\mathbf{R}_\mathrm{y}}
\newcommand{\Ryest}{\hat{\mathbf{R}}_\mathrm{y}}
\newcommand{\Rx}{\mathbf{R}_\mathrm{x}}
\newcommand{\Rxest}{\hat{\mathbf{R}}_\mathrm{x}}
\newcommand{\Amat}{\mathbf{A}}
\newcommand{\Bmat}{\mathbf{B}}
\newcommand{\phix}[1]{\phi_{\mathrm{x},#1}}
\newcommand{\phin}[1]{\phi_{\mathrm{n},#1}}
\newcommand{\ExOp}[1]{\mathcal{E}\{#1\}}
\newcommand{\SNR}{\mathrm{SNR}}
\newcommand{\evec}{\mathbf{e}}
\newcommand{\w}{\mathbf{w}}
\newcommand{\Smat}{\mathbf{S}}

\copyrightnotice{978-1-6654-6867-1/22/\$31.00~\copyright2022 IEEE}
                   
\title{Bias Analysis of Spatial Coherence-Based RTF Vector Estimation\\ for Acoustic Sensor Networks in a Diffuse Sound Field}
\name{Wiebke Middelberg and Simon Doclo\thanks{This work was funded by the Deutsche Forschungsgemeinschaft (DFG, German Research Foundation) under Germany's Excellence Strategy - EXC 2177/1 - Project ID 390895286 and Project ID 352015383 - SFB 1330 B2.}}
\address{University of Oldenburg, Department of Medical Physics and Acoustics\\ and Cluster of Excellence Hearing4all, Oldenburg, Germany}
\begin{document}
\ninept
\maketitle
\begin{abstract}
In many multi-microphone algorithms, an estimate of the relative transfer functions (RTFs) of the desired speaker is required. Recently, a computationally efficient RTF vector estimation method was proposed for acoustic sensor networks, assuming that the spatial coherence (SC) of the noise component between a local microphone array and multiple external microphones is low.
Aiming at optimizing the output signal-to-noise ratio (SNR), this method linearly combines multiple RTF vector estimates, where the complex-valued weights are computed using a generalized eigenvalue decomposition (GEVD).
In this paper, we perform a theoretical bias analysis for the SC-based RTF vector estimation method with multiple external microphones. Assuming a certain model for the noise field, we derive an analytical expression for the weights, showing that the optimal model-based weights are real-valued and only depend on the input SNR in the external microphones.
Simulations with real-world recordings show a good accordance of the GEVD-based and the model-based weights. Nevertheless, the results also indicate that in practice, estimation errors occur which the model-based weights cannot account for.

\end{abstract}
\begin{keywords}
RTF vector estimation, MVDR beamforming, acoustic sensor networks, bias analysis, external microphones
\end{keywords}
\section{Introduction}
\label{sec:intro}

In many speech communication applications, such as hearing aids, conferencing and hands-free systems, speech intelligibility is often decreased by the presence of undesired noise sources. Therefore, noise reduction is essential to increase the speech intelligibility \cite{Hamacher2008, Doclo2015}.
A well-known multi-microphone noise reduction algorithm is the minimum variance distortionless response (MVDR) beamformer \cite{VanVeen1988a,Doclo2010}. It has been shown that the MVDR beamformer can be steered without requiring the microphone positions to be known by the relative transfer function (RTF) vector of the desired speaker \cite{Gannot2001,Gannot2017}. The RTFs relate the speech component in a reference microphone signal to the speech component in all other microphone signals.

Instead of using a compact microphone array with closely spaced\linebreak microphones for noise reduction, there has been a recent trend towards using spatially distributed microphones, also referred to as acoustic\linebreak sensor networks, where the positions of the microphones are typically not known \cite{Yee2017,Ali2019,Goessling2018_itg_b,Corey2021,Middelberg2021,Bertrand2011}.
For an acoustic sensor network consisting of a local microphone array (LMA), e.g., binaural hearing aids, and one additional external microphone, a computationally efficient RTF vector estimation method was proposed \cite{Goessling2018_iwaenc_a,GoesslingICASSP2019}, assuming that the spatial coherence (SC) between the noise component in the external microphone and the LMA is low. A bias analysis in \cite{GoesslingICASSP2019} showed that the RTF vector entry corresponding to the external microphone has a real-valued bias (i.e., a systematic\linebreak estimation error), which increases with decreasing input signal-to-noise ratio (SNR) in the external microphone.
In \cite{GoesslingWASPAA2019}, the SC RTF vector estimation method was extended to exploit multiple external microphones. In the so-called mSNR approach, it was proposed to linearly combine different (biased) RTF vector estimates in order to maximize the output SNR of an MVDR beamformer. It was shown that the optimal combination weights can be computed via a generalized eigenvalue decomposition (GEVD).

As an extension of the bias analysis for the SC method with one external microphone in \cite{GoesslingICASSP2019}, in this paper we perform a bias analysis for the mSNR approach with multiple external microphones in a diffuse noise field.
Using a model of the diffuse sound field, we derive an analytic expression for the optimal combination weights. The model-based weights are found to be real-valued and only depend on the input SNRs in the external microphones. In addition, we show that the biases of the external microphone entries in the RTF vector estimate obtained using the mSNR approach are smaller than the corresponding biases in the RTF vector estimates obtained using the SC method. An experimental evaluation with real-world recordings shows that the real-valued model-based weights and the real value of the generally complex-valued GEVD-based weights are similar. Nevertheless, although the model-based weights are useful for analyzing the bias of the RTF vector entries, simulation results show that in terms of SNR improvement the GEVD-based weights outperform the model-based weights when used in an MVDR beamformer.

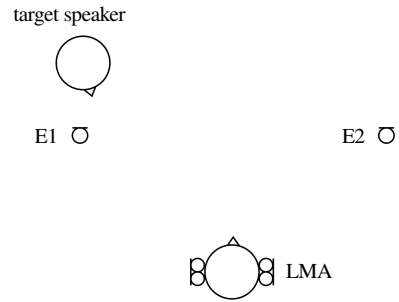
\begin{figure}[tb]
	\centerline{\includegraphics[width=0.53\linewidth]{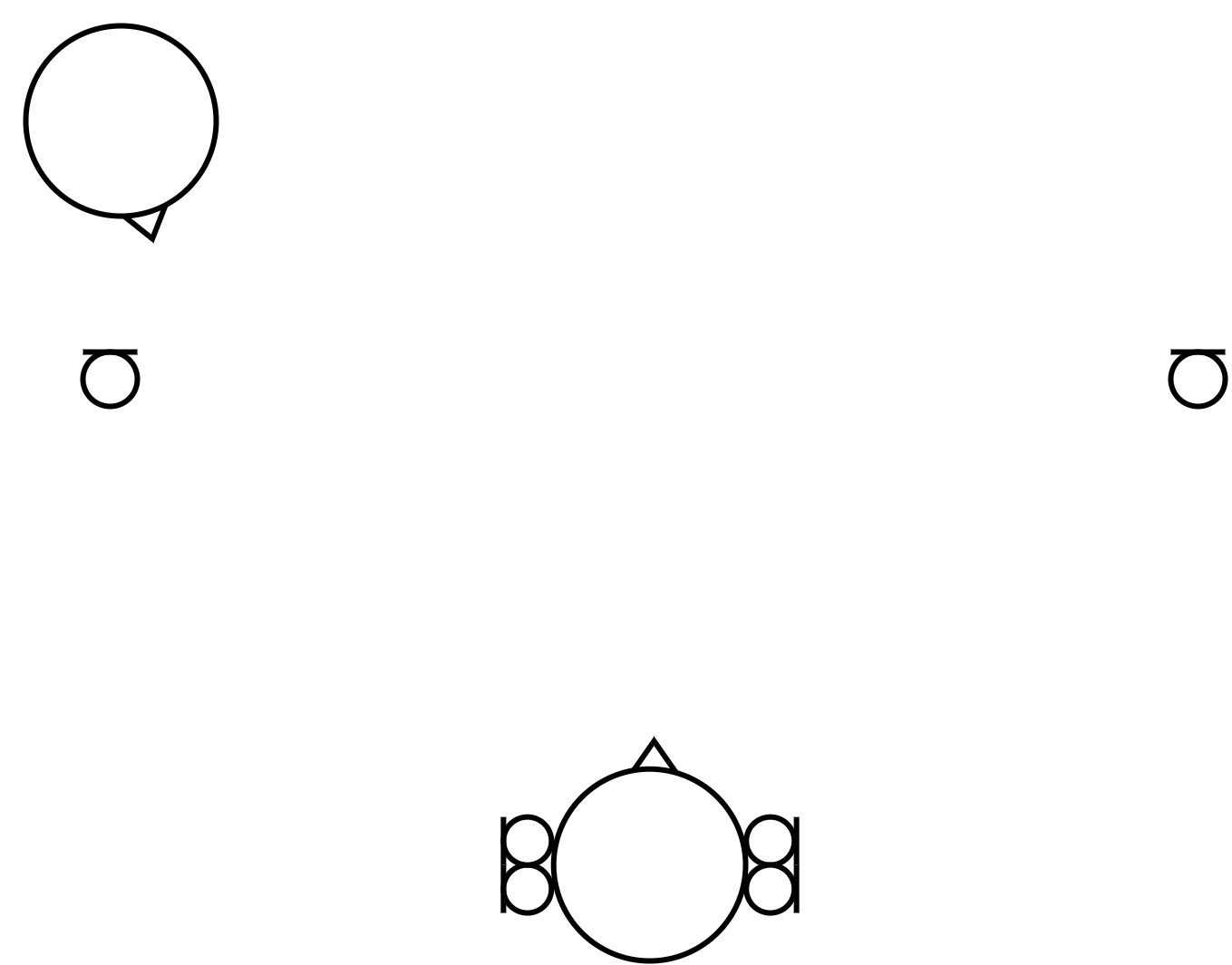}
	\put(-143,105){\footnotesize target speaker}
	\put(-41,8){\footnotesize LMA}
	\put(-135,59){\footnotesize E1}
	\put(-20,59){\footnotesize E2}
	}
	\vspace{-0.3cm}
	\caption{Acoustic scene with a target speaker, binaural hearing aids (as the LMA) mounted on a dummy head and two external microphones (E1 and E2).}
	\label{fig:config}
	\vspace{-0.5cm}
\end{figure}

\section{Signal Model and Notation}
\label{sec:SignalModel}

We consider an acoustic sensor network consisting of an LMA with $\Ma$ microphones and $\Me$ external microphones, depicted in Fig. \ref{fig:config}, giving a total of $M = \Ma+\Me$ microphones. In the short-time Fourier transform (STFT)-domain, the noisy $m$-th microphone signal is defined as 
\begin{equation}\label{eq:Ystft}
Y_m(k,l) = X_m(k,l) + N_m(k,l)\, ,\quad m \in \{1,\dots,M\}\; ,
\end{equation}
where $X_m(k,l)$ and $N_m(k,l)$ denote the speech and noise component, respectively. For brevity, the frequency bin index $k$ and the frame index $l$ are omitted in the remainder of this paper.
Assuming a multiplicative transfer function model, the speech component can be written as $X_m=H_mX_1$, where $H_m$ denotes the RTF of the target speaker between the $m$-th microphone and the reference microphone (chosen as the first microphone without loss of generality).
The $M$-dimensional signal vector $\y = [Y_{\asub,1}, Y_{\asub,2}, \dots, Y_{\asub,{\Ma}},Y_{\esub,1},Y_{\esub,2},\dots,Y_{\esub,{\Me}}]^T$
contains both the LMA as well as the external microphone signals, where $\{\cdot\}^T$ denotes the transpose operator.
The signal vector can be written as
\begin{equation}\label{eq:addsignals}
    \y\; =\; \x +\n \;=\; \h X_1 + \n\; ,
\end{equation}
where the speech and noise vector $\x$ and $\n$ are defined similarly to $\y$ and $\h$ denotes the $M$-dimensional RTF vector of the target speaker, with $H_1 = 1$.

Assuming speech and noise to be uncorrelated, the noisy covariance matrix $\Ry$ is equal to
\begin{equation}\label{eq:CovMat}
    \Ry = \ExOp{\y\y^H} = \ExOp{\x\x^H} +\ExOp{\n\n^H} = \Rx + \Rn\, ,
\end{equation}
where $\ExOp{\cdot}$ denotes the expectation operator, $\{\cdot\}^H$ denotes the Hermitian transpose operator and $\Rx$ and $\Rn$ are the speech and noise covariance matrices, respectively.
Based on \eqref{eq:addsignals}, the speech covariance matrix is a rank-1 matrix, spanned by the RTF vector, i.e., 
\begin{equation}\label{eq:rank1}
    \Rx = \phix{1}\h\h^H\, ,
\end{equation}
where $\phix{m} = \ExOp{|X_m|^2}$ denotes the speech power spectral density (PSD) in the $m$-th microphone.
In the remainder of this paper, we will assume that the noise component in any external microphones is uncorrelated with the noise component in any other microphone, i.e.,
\begin{equation}\label{eq:rn0}
    \rn{i,j} = 0\; ,\quad \text{if}\; (i\in \{\Ma+1,\dots,M\}\lor j \in \{\Ma+1,\dots,M\})\, \land\, i\neq j\, .
\end{equation}
Although this assumption obviously does not hold for every acoustic scenario (especially not for coherent interfering sources), this assumption holds well for a diffuse noise field (e.g., babble noise) where all external microphones are placed at a large distance to all other microphones.

The MVDR beamformer minimizing the output noise PSD while preserving the speech component in the reference microphone signal is given by \cite{Doclo2015,Gannot2017}
\begin{equation}\label{eq:MVDR}
    \w = \frac{\Rninv\h}{\h^H\Rninv\h}\, ,
\end{equation}
where the beamformer output signal is equal to $Z=\w^H\y$. The input SNR of the $m_\esub$-th external microphone signal and the biased SNR of the beamformer output signal are given by
\begin{equation}\label{eq:SNRdef}
    \SNR_{\esub,m_\esub} = \frac{\phix{\esub,m_\esub}}{\phin{\esub,m_\esub}}\; ,
\end{equation}
\begin{equation}\label{eq:SNRout}
    \SNR_\text{out}^{\rm b} = \frac{\w^H\Ry\w}{\w^H\Rn\w} = \frac{\w^H\Rx\w}{\w^H\Rn\w} + 1 = \SNR_{\rm out} +1\; ,
\end{equation}
where $\phix{\esub,m_\esub}$ and $\phin{\esub,m_\esub}$ denote the speech and noise PSD in the $m_\esub$-th external microphone, respectively, and $\SNR_\text{out}$ denotes the unbiased output SNR.

\section{RTF Estimation Using External Microphones}
\label{sec:RTF}
In this section, we present the SC-based RTF vector estimation method for one external microphone proposed in \cite{GoesslingICASSP2019} and its extension for multiple external microphones \cite{GoesslingWASPAA2019}.
Exploiting the assumption about spatial correlation of the noise field in \eqref{eq:rn0}, $\Me$ different estimates of the RTF vector $\h$ can be efficiently obtained by selecting and normalizing the column of the noisy covariance matrix corresponding to the external microphone \cite{GoesslingICASSP2019}, i.e.,
\begin{equation}\label{eq:RTF_SC}
    \hestSC_{m_\esub} = \frac{\Ry\evec_{\esub,m_\esub}}{\evec_1^T\Ry\evec_{\esub,m_\esub}}\, ,\quad m_\esub \in \{1,\dots,\Me\}\, ,
\end{equation}
where $\evec_{\esub,m_\esub}$ is an $M$-dimensional selection vector for the $m_\esub$-th external microphone.
Using the model for $\Rx$ and $\Rn$ in \eqref{eq:rank1} and \eqref{eq:rn0}, it can be easily shown that all entries of the RTF vector estimate in \eqref{eq:RTF_SC} are unbiased, except for the entry corresponding to the $m_\esub$-th external microphone, which is systematically biased and equal to \cite{GoesslingICASSP2019}
\begin{equation}\label{eq:biasSingle}
    \boxed{\tilde{H}_{\esub,m_\esub}^{\rm SC}=\evec_{\esub,m_\esub}^T\hest_{m_\esub}^{\rm SC} = \left(1+\frac{1}{\SNR_{\esub,m_\esub}}\right) H_{\esub,m_\esub} }
\end{equation}
This means that the estimated RTF is equal to the true RTF multiplied with a bias term which directly depends on the input SNR in the used $m_\esub$-th external microphone.

To decrease the biases of the estimated external RTFs, it was proposed in \cite{GoesslingWASPAA2019} to linearly combine the $\Me$ RTF vector estimates in \eqref{eq:RTF_SC} i.e.,
\begin{equation}\label{eq:h_comb}
    \hestMAX = \Hest\avec\, ,
\end{equation}
\begin{equation}\label{eq:Hest}
    \Hest = [\hest_1^{\rm SC}, \hest_2^{\rm SC},\dots,\hest_{\Me}^{\rm SC}]\, ,
\end{equation}
where the elements of the $\Me$-dimensional (complex-valued) weights vector $\avec$ are constrained to sum up to 1, such that $\tilde{H}_1 = 1$.
In the mSNR approach in \cite{GoesslingWASPAA2019}, the optimal weight vector is calculated by maximizing the (biased) output SNR of the RTF-steered MVDR beamformer, i.e.,
\begin{equation}\label{eq:optSNR}
    \max_{\avec}\; \SNR_{\rm out}^{\rm b}(\avec)\, , \quad \text{s.t.}\quad \ones^T\avec = 1\, ,
\end{equation}
where $\ones$ is an $\Me$-dimensional vector of ones. By using the MVDR beamformer in \eqref{eq:MVDR} with the RTF vector estimate in \eqref{eq:h_comb} in \eqref{eq:SNRout}, the cost function $J(\avec)$ corresponding to the biased output SNR can be written as
\begin{equation}\label{eq:CostFunc}
    J(\avec) = \SNR_{\rm out}^{\rm b}(\avec) = \frac{\avec^H\Amat\avec}{\avec^H\Bmat\avec}\, ,
\end{equation}
where the $\Me\times\Me$-dimensional matrices $\Amat$ and $\Bmat$ are defined as\linebreak $\Amat = \Hest^H\Rninv\Ry\Rninv\Hest$ and $\Bmat = \Hest^H\Rninv\Hest$ in \cite{GoesslingWASPAA2019}. 
The optimal weighting vector maximizing \eqref{eq:CostFunc} can hence be calculated using the GEVD of $\Amat$ and $\Bmat$, where the normalization constraint in \eqref{eq:optSNR} is applied subsequently, i.e.,
\begin{equation}\label{eq:GEVD}
    \avec^{\rm GEVD} = \frac{\mathcal{P}\{\Bmat^{-1}\Amat\}}{\ones^T\mathcal{P}\{\Bmat^{-1}\Amat\}}\, ,
\end{equation}
where $\mathcal{P}\{\cdot\}$ denotes the principal eigenvector operator.

\section{Theoretical Bias Analysis}
\label{sec:Bias}

In this section, we investigate the biases of the RTF vector estimate $\hestMAX$ in \eqref{eq:h_comb} obtained by combining the RTF vector estimates $\hestSC_{m_\esub}$, with $m_\esub \in \{1,...,\Me\}$, in \eqref{eq:RTF_SC} using the mSNR approach, i.e., using the weight vector in \eqref{eq:GEVD}.
To that end, we write the matrix of estimated RTF vector estimates $\Hest$ in \eqref{eq:Hest} in terms of the true RTF vector $\h$ and a bias matrix $\Emat$, i.e., 
\begin{equation}\label{eq:Hestmod}
    \Hest = \Hmat+\Emat\; ,
\end{equation}
where, using \eqref{eq:RTF_SC}, $\Hmat = \h\ones^T$ and 
\begin{equation}\label{eq:Emat}
    \Emat = \left[\begin{array}{cccc} 
    \zeros{\Ma}{1} & \zeros{\Ma}{1} & \dots & \zeros{\Ma}{1}\\
    \frac{H_{\esub,1}}{\SNR_{\esub,1}} & 0 & \dots & 0\\
    0 & \frac{H_{\esub,2}}{\SNR_{\esub,2}} & \dots & 0\\
    \vdots & \vdots & \ddots & \vdots\\
    0 & \dots & \dots & \frac{H_{\esub,\Me}}{\SNR_{\esub,\Me}}
    \end{array}\right]\, .
\end{equation}
By substituting $\Hest$ in \eqref{eq:Hestmod} into the cost function $J(\avec)$ in \eqref{eq:optSNR}, we obtain
\begin{equation}\label{eq:JaMod}
    J(\avec) = \frac{\avec^H(\Amat_1 + \Amat_2 + \Amat_2^H + \Amat_3)\avec}{\avec^H(\Bmat_1 + \Bmat_2 + \Bmat_2^H + \Bmat_3)\avec}\; ,
\end{equation}
with
\begin{equation}\label{eq:Amat1Bmat3}
\begin{aligned}
    &\Amat_1 = \ones\;(\h^H\Rninv\Ry\Rninv\h)\;\ones^T\, ,\\
    &\Amat_2 = \ones\h^H\Rninv\Ry\Rninv\Emat\; ,\\
    &\Amat_3 = \Emat^H\Rninv\Ry\Rninv\Emat\; ,
\end{aligned}
\quad
\begin{aligned}
    &\Bmat_1 = \ones\;(\h^H\Rninv\h)\;\ones^T\; , \\
    &\Bmat_2 = \ones\h^H\Rninv\Emat\; , \\
    &\Bmat_3 = \Emat^H\Rninv\Emat\; .
\end{aligned}
\end{equation}
We now aim at simplifying \eqref{eq:JaMod} by carefully investigating all matrices in \eqref{eq:Amat1Bmat3}. The matrices $\Amat_1$ and $\Bmat_1$ can be written as a scaled matrix of ones, i.e., $\Amat_1 = (\h^H\Rninv\Ry\Rninv\h)\;\ones\ones^T = a_1 \ones\ones^T$ and $\Bmat_1=(\h^H\Rninv\h)\;\ones\ones^T= b_1 \ones\ones^T$, where $a_1$ and $b_1$ are real-valued positive scalars as they can be regarded as the quadratic form of a positive definite matrix.\\
Since the matrices $\Amat_2$, $\Amat_3$, $\Bmat_2$ and $\Bmat_3$ all contain $\Rninv\Emat$, we consider this term in more detail.
Using \eqref{eq:rn0} and \eqref{eq:Emat}, $\Rninv\Emat$ is equal to
\begin{equation}\label{eq:RninvE}
    \Rninv\Emat = \left[\begin{array}{ccc} 
    \zeros{\Ma}{1} &  \dots & \zeros{\Ma}{1}\\
    \frac{H_{\esub,1}}{\SNR_{\esub,1}}\frac{1}{\phin{\esub,1}}  & \dots & 0\\
    \vdots & \ddots & \vdots\\
    0  & \dots & \frac{H_{\esub,\Me}}{\SNR_{\esub,\Me}}\frac{1}{\phin{\esub,\Me}}
    \end{array}\right]\, .
\end{equation}
By realizing that $|H_{\esub,m_\esub}|^2 = \phix{\esub,m_\esub}/\phix{1}$, the left-hand multiplication of \eqref{eq:RninvE} with $\h^H$ yields
\begin{equation}\label{eq:hRninvE}
    \h^H\Rninv\Emat = (1/\phix{1})\;\ones^T\, .
\end{equation}
Hence, the matrix $\Bmat_2$ can also be written as a scaled matrix of ones, i.e., $\Bmat_2 = b_2\,\ones\ones^T$ with $b_2 = 1/\phix{1}$ a real-valued positive scalar.
Using \eqref{eq:CovMat}, \eqref{eq:rank1} and the simplification in \eqref{eq:hRninvE}, the matrix $\Amat_2$ is equal to
\begin{equation}\label{eq:Amat2trafo}
\begin{split}
    \Amat_2 &= \ones\h^H\Rninv\phix{1}\h\h^H\Rninv\Emat \;+\; \ones\h^H\Rninv\Emat\\
        &= (\h^H\Rninv\h)\;\ones\ones^T\; +\; (1/\phix{1})\;\ones\ones^T\\
                    &= (b_1+b_2)\;\ones\ones^T\, ,
\end{split}
\end{equation}
i.e., $\Amat_2$ can also be written as a scaled matrix of ones.
Finally, using \eqref{eq:Emat} and \eqref{eq:RninvE}, the matrix $\Bmat_3$ can be written as
\begin{equation}\label{eq:Bmat3trafo}
    \Bmat_3 = \Emat^H\Rninv\Emat = \frac{1}{\phix{1}}\underbrace{\left[\begin{array}{ccc}
                \frac{1}{\SNR_{\esub,1}} & \dots &  0\\
                \vdots & \ddots & \vdots\\
                0 & \dots & \frac{1}{\SNR_{\esub,\Me}}
            \end{array}\right]}_{\Smat}\, ,
\end{equation}
where $\Smat$ is a full-rank matrix containing the inverse input SNR in the external microphones on its diagonal.
Using \eqref{eq:CovMat}, \eqref{eq:rank1} and \eqref{eq:hRninvE}, the matrix $\Amat_3$ can be written as
\begin{equation}\label{eq:Amat3trafo}
\begin{split}
    \Amat_3 &= \Emat^H\Rninv\phix{1}\h\h^H\Rninv\Emat + \Emat^H\Rninv\Emat\\
            &= (1/\phix{1})\;\ones\ones^T + (1/\phix{1})\;\Smat\\
            &= b_2\;\ones\ones^T + b_2\;\Smat\; .
\end{split}
\end{equation}
By now considering the normalization constraint in \eqref{eq:optSNR}, it becomes evident that $\avec^H\ones\ones^T\avec = 1$. 
Together with the derived matrix expressions in the previous paragraph, the cost function in \eqref{eq:JaMod} then reduces to
\begin{equation}\label{eq:Ja2}
    J(\avec) = \frac{a + c(\avec)}{b + c(\avec)}\, ,
\end{equation}
where $a=a_1+2(b_1+b_2)+b_2$, $b = b_1+2b_2$ and
\begin{equation}
    c(\avec)=b_2\;\avec^H\Smat\avec\, ,
\end{equation}
with $\Smat$ defined in \eqref{eq:Bmat3trafo}.
Since $a\ge b$ and $c(\avec)$ is real-valued and positive, maximizing $J(\avec)$ in \eqref{eq:Ja2} corresponds to minimizing $c(\avec)$. Hence, the optimal weights are the solution of the constraint optimization problem
\begin{equation}\label{eq:optc}
    \min_{\avec}\; \avec^H\Smat\avec\, , \quad \text{s.t.}\quad \ones^T\avec = 1\, ,
\end{equation}
i.e.,
\begin{equation}\label{eq:avSinv}
    \avec^{\rm model} = \frac{\Smat^{-1}\ones}{\ones^T\Smat^{-1}\ones}\; .
\end{equation}
By using \eqref{eq:Bmat3trafo}, these weights can be written as
\begin{equation}\label{eq:avfinal}
    \boxed{\avec^{\rm model} = \frac{1}{\sum_{m_\esub'=1}^{\Me} \SNR_{\esub,m_\esub'}}\left[\begin{array}{c}
        \SNR_{\esub,1}   \\
          \vdots\\
        \SNR_{\esub,\Me}  
    \end{array}\right]}
\end{equation}
Hence, if the assumed model for the speech and noise covariance matrix $\Rx$ and $\Rn$ in \eqref{eq:rank1} and \eqref{eq:rn0} holds, the optimal weights for the mSNR approach are real-valued and equal to the normalized input SNRs in the external microphones.
This means that in the linear combination for computing $\hestMAX$ in \eqref{eq:h_comb}, SC RTF vector estimates corresponding to external microphones with a higher input SNR are assigned a larger weight than those with a low input SNR, which seems very intuitive.

When analyzing the bias of the RTF vector estimates $\hestMAX$ in \eqref{eq:h_comb} using the model weights in \eqref{eq:avfinal}, the first $\Ma$ entries (corresponding to the LMA) are obviously still unbiased while the bias of the last $\Me$ entries (corresponding to the external microphones) is equal to, using \eqref{eq:biasSingle},
\begin{equation}\label{eq:biasFinal}
    \boxed{\tilde{H}_{\esub,m_\esub}^{\rm mSNR} = \left(1+\frac{1}{\sum_{m_\esub'=1}^{\Me}\SNR_{\esub,m_\esub'}}\right) H_{\esub,m_\esub}}
\end{equation}
In contrast to the individual RTF vector estimates in \eqref{eq:biasSingle}, it can be observed that the bias of the optimal linear combination is the same for all entries corresponding to the external microphones. In addition, the bias in \eqref{eq:biasFinal} for the mSNR approach (exploiting all external microphones) is always smaller than the bias in \eqref{eq:biasSingle} for the individual RTF vector estimates.

\section{Evaluation}
\label{sec:Eval}

The following evaluation aims at demonstrating accordance and deviations between the theoretically found weights of the mSNR approach (referred to as "model") in \eqref{eq:avfinal} and the solution using a GEVD of the respective covariance matrices (referred to as "GEVD") in \eqref{eq:GEVD}. It should be noted that in theory, when all assumptions are fulfilled, the GEVD- and model-based weights are identical. However, the assumptions used in the bias analysis in Section \ref{sec:Bias}, such as the statistical independence of speech and noise in \eqref{eq:CovMat}, the rank-1 model in \eqref{eq:rank1} and the spatial coherence assumption in \eqref{eq:rn0}, do not perfectly hold in practice, where estimates of the covariance matrices are used.

\subsection{Setup and Implementation}
\label{subsec:Setup}


The evaluation was performed using real-world signals that were recorded in a laboratory room with a size of about (7 × 6 × 2.7) $\rm m^3$ and a reverberation time of approximately 400 ms. For the LMA, binaural hearing aids with 2 microphones per side ($\Ma$ = 4) with an inter-microphone distance of about 7 mm were mounted on a KEMAR dummy head. In addition, $\Me$ = 2 external microphones were placed to the left and the right front of the dummy head, both at a distance of about 2.3 m from the dummy head, as depicted in Fig. \ref{fig:config}.
The desired speech source was a male German speaker, walking from the left to the right side of the dummy head, i.e., from the first external microphone (E1) to the second external microphone (E2).
Pseudo-diffuse background noise was generated using four loudspeakers facing the corners of the laboratory, playing back different multi-talker recordings.
The speech and noise components were recorded separately and were subsequently mixed. The broadband input SNR in the LMA microphones varied from about 0 to 6 dB due to the movement of the speech source, while the input SNR in the external microphone signals varied as shown in the upper panel of Fig. \ref{fig:weights}. The sampling rate for all signals was 16 kHz.

All processing was performed in the STFT-domain with a frame length of 32 ms, a square-root-Hann window for analysis and synthesis and an overlap of 50\%.
To allow for a good visualization of the theoretical findings of the bias analysis, oracle estimates of the covariance matrices $\Ry$, $\Rx$ and $\Rn$ were computed on the separate signal components $\y$, $\x$ and $\n$. To account for the dynamic acoustic scenario, the covariance matrices were updated by recursive smoothing with time constants of 250 ms for $\Ry$ and $\Rx$ and 1 s for $\Rn$, respectively, where $\Ry$ and $\Rx$ were only updated if speech was active (determined via an oracle broadband voice activity detection).

As a performance measure, we considered the broadband SNR improvement ($\Delta\SNR = \SNR_{\rm out}-\SNR_{\rm in}$) of the RTF-steered MVDR beamformer using
\begin{enumerate}
    \item the SC RTF vector estimates $\hestSC_{m_\esub}$ in \eqref{eq:RTF_SC} from the individual external microphones (SC-1 and SC-2), computed using the estimated covariance matrix $\Ryest$.
    \item the mSNR combination $\hestMAX$ in \eqref{eq:h_comb} using the (complex-valued) GEVD-based weights $\avec^{\rm GEVD}$ in \eqref{eq:GEVD}, computed using the estimated covariance matrices $\Ryest$ and $\Rnest$.
    \item the mSNR combination $\hestMAX$ in \eqref{eq:h_comb} using the (real-valued) model-based weights in \eqref{eq:avfinal}, computed using the estimated covariance matrices $\Rxest$ and $\Rnest$.
\end{enumerate}
The SNR improvement was averaged over the left and right side of the hearing aids to reduce effects caused by the moving speaker. Input and output SNR were computed in the time-domain using the shadow filter approach, only when the speaker was active.


\subsection{Results}
\label{subsec:Results}

\begin{figure}
    \centering
    \includegraphics[width=\linewidth]{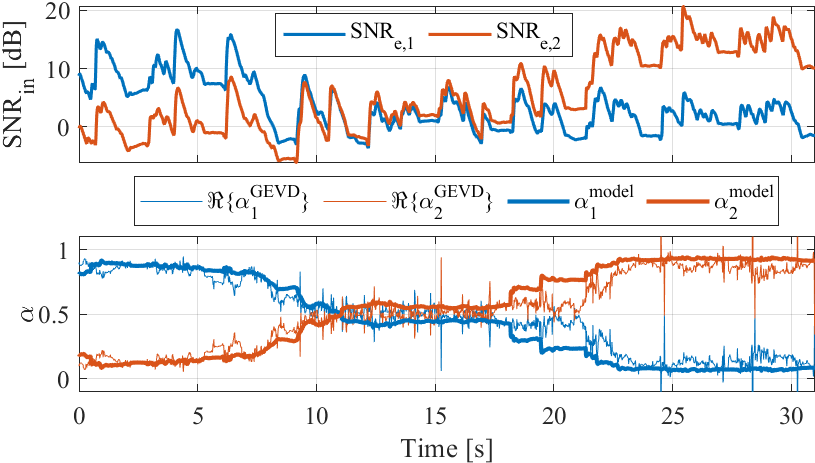} 
    \vspace{-0.6cm}
    \caption{Input SNR in both external microphones (upper panel) and weights of RTF vector estimates obtained by the GEVD-based and the model-based mSNR combination (lower panel).}
    \label{fig:weights}
\end{figure}

The upper panel of Fig. \ref{fig:weights} shows the input SNR in the two external microphones (averaged over frequency) plotted over time. As expected from the movement of the source from E1 to E2, the input SNR decreases in E1 while increasing in E2 over time. The GEVD- and model-based weights to combine the RTF vector estimated obtained are shown in the lower panel of Fig. \ref{fig:weights}. For the model-based weights $\avec^{\rm model}$ it can be observed that the weight of the RTF vector estimate obtained from E1 $\alpha_1^{\rm model}$ (thick blue line) initially is almost equal to 1, meaning that this estimate dominates in the linear combination. Over time, with decreasing input SNR in E1, the weight $\alpha_1^{\rm model}$ also decreases, whereas the weight of the RTF vector estimate obtained from E2 $\alpha_2^{\rm model}$ (thick red line) increases. For the GEVD-based weights $\alpha_1^{\rm GEVD}$ and $\alpha_2^{\rm GEVD}$ a similar behavior can be observed (thin lines), showing a good accordance between the theoretical model-based and the practical GEVD-based weights.
Nevertheless, since the GEVD-based weights are complex-valued and we only investigate the real part, there exists a larger deviation between practice and theory than can be observed from this figure.

\begin{figure}
    \centering
    \includegraphics[width=\linewidth]{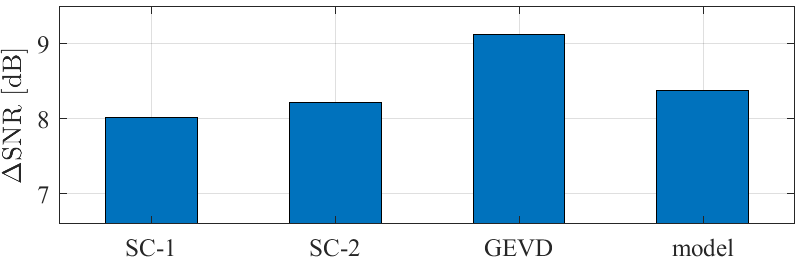} 
    \vspace{-0.7cm}
    \caption{Time-average broadband output SNR for the single SC RTF vector estimates (SC-1 and SC-2) and combinations of estimates using the GEVD- and the model-based weights used in an MVDR beamformer.}
    \label{fig:SNR_av}
    \vspace{-0.3cm}
\end{figure}

Fig. \ref{fig:SNR_av} shows the SNR improvement (averaged over time) of the considered RTF-steered MVDR beamformers. It can be seen that both SC RTF vector estimates (SC-1 and SC-2) perform very similarly, yielding an SNR improvement of about 8 dB.
The model-based mSNR combination leads to a slightly better SNR improvement than the individual RTF vector estimates, SC-1 and SC-2. 
The GEVD-based mSNR combination clearly yields the best performance, leading to an SNR improvement of more than 9 dB. Hence, despite (the real parts of) the GEVD- and the model-based weights being rather similar, the complex-valued GEVD-based weights lead to better results than the real-valued model-based weights.
The deviations of the theoretically identical GEVD- and model-based weights can be explained by the model assumptions, such as the rank-1 model in \eqref{eq:rank1} and the spatial coherence assumption about the noise field in \eqref{eq:rn0}, not perfectly holding in practice.
This leads to generally complex-valued GEVD-based weights, which can - to some extent - compensate for these deviations and estimation errors.

\section{Conclusions}
\label{sec:Conclusion}

In this paper, we performed a theoretical analysis of the mSNR approach, which linearly combines (biased) RTF vector estimates obtained using the SC method.
We derived an analytic expression for the optimal weights of the linear combination (usually obtained by a complex-valued GEVD), which is real-valued and only depends on the input SNRs in the external microphones.
Furthermore, we showed that the biases of the optimal combination are smaller than the bias of every individual RTF vector estimate.
In a simulation with real-world recordings, we showed that the theoretically equivalent GEVD- and model-based weights strongly resemble each other but some deviations can be observed in practice, leading to a better performance of the complex-valued GEVD-based weights in terms of SNR improvement.
The observed differences between the model- and GEVD-based weights can be explained by the model assumptions made in the analysis not perfectly holding in practice.


\balance 
\bibliographystyle{IEEEbib}
\bibliography{bib}

\begin{thebibliography}{10}

\bibitem{Hamacher2008}
V.~Hamacher, U.~Kornagel, T.~Lotter, and H.~Puder,
\newblock ``{Binaural signal processing in hearing aids: {T}echnologies and
  algorithms},''
\newblock in {\em Advances in Digital Speech Transmission}, chapter~14, pp.
  401--429. Wiley, 2008.

\bibitem{Doclo2015}
S.~Doclo, W.~Kellermann, S.~Makino, and S.~E. Nordholm,
\newblock ``{Multichannel signal enhancement algorithms for assisted listening
  devices: Exploiting spatial diversity using multiple microphones},''
\newblock {\em IEEE Signal Processing Magazine}, vol. 32, no. 2, pp. 18--30,
  Mar. 2015.

\bibitem{VanVeen1988a}
B.~D. {Van Veen} and K.~M. Buckley,
\newblock ``{Beamforming: {A} versatile approach to spatial filtering},''
\newblock {\em IEEE ASSP Magazine}, vol. 5, no. 2, pp. 4--24, Apr. 1988.

\bibitem{Doclo2010}
S.~Doclo, S.~Gannot, M.~Moonen, and A.~Spriet,
\newblock ``Acoustic beamforming for hearing aid applications,''
\newblock in {\em Handbook on Array Processing and Sensor Networks}, pp.
  269--302. Wiley, 2010.

\bibitem{Gannot2001}
S.~Gannot, D.~Burshtein, and E.~Weinstein,
\newblock ``{Signal enhancement using beamforming and nonstationarity with
  applications to speech},''
\newblock {\em IEEE Trans. on Signal Processing}, vol. 49, no. 8, pp.
  1614--1626, Aug. 2001.

\bibitem{Gannot2017}
S.~Gannot, E.~Vincent, S.~Markovich-Golan, and A.~Ozerov,
\newblock ``A consolidated perspective on multi-microphone speech enhancement
  and source separation,''
\newblock {\em IEEE/ACM Trans. on Audio, Speech, and Language Processing}, vol.
  25, pp. 692--730, Apr. 2017.

\bibitem{Yee2017}
D.~Yee, H.~Kamkar-Parsi, R.~Martin, and H.~Puder,
\newblock ``{A noise reduction post-filter for binaurally-linked
  single-microphone hearing aids utilizing a nearby external microphone},''
\newblock {\em IEEE/ACM Trans. on Audio, Speech, and Language Processing}, vol.
  26, no. 1, pp. 5--18, Jul. 2017.

\bibitem{Ali2019}
R.~Ali, G.~Bernardi, T.~van Waterschoot, and M.~Moonen,
\newblock ``Methods of extending a generalized sidelobe canceller with external
  microphones,''
\newblock {\em IEEE/ACM Trans. on Audio, Speech, and Language Processing}, vol.
  27, pp. 1349--1364, Sep. 2019.

\bibitem{Goessling2018_itg_b}
N.~G{\"{o}}{\ss}ling and S.~Doclo,
\newblock ``{{RTF}-based binaural {MVDR} beamformer exploiting an external
  microphone in a diffuse noise field},''
\newblock in {\em Proc. ITG Conference on Speech Communication}, Oldenburg,
  Germany, Oct. 2018, pp. 106--110.

\bibitem{Corey2021}
R.~M. Corey and A.~C. Singer,
\newblock ``Adaptive binaural filtering for a multiple-talker listening system
  using remote and on-ear microphones,''
\newblock in {\em Proc. IEEE Workshop on Applications of Signal Processing to
  Audio and Acoustics}, New Paltz, USA, Oct. 2021, pp. 1--5.

\bibitem{Middelberg2021}
W.~Middelberg and S.~Doclo,
\newblock ``{Comparison of Generalized Sidelobe Canceller Structures
  Incorporating External Microphones for Joint Noise and Interferer
  Reduction},''
\newblock in {\em Proc. ITG Conference on Speech Communication}, Kiel, Germany,
  Oct. 2021, pp. 104--108.

\bibitem{Bertrand2011}
A.~Bertrand,
\newblock ``Applications and trends in wireless acoustic sensor networks: A
  signal processing perspective,''
\newblock in {\em 18th IEEE Symposium on Communications and Vehicular
  Technology in the Benelux (SCVT)}, Nov. 2011, pp. 1--6.

\bibitem{Goessling2018_iwaenc_a}
N.~G{\"{o}}{\ss}ling and S.~Doclo,
\newblock ``{Relative transfer function estimation exploiting spatially
  separated microphones in a diffuse noise field},''
\newblock in {\em Proc. International Workshop on Acoustic Signal Enhancement},
  Tokyo, Japan, Sep. 2018, pp. 146--150.

\bibitem{GoesslingICASSP2019}
N.~G{\"o}{\ss}ling and S.~Doclo,
\newblock ``{RTF}-steered binaural {MVDR} beamforming incorporating an external
  microphone for dynamic acoustic scenarios,''
\newblock in {\em Proc. IEEE International Conference on Acoustics, Speech and
  Signal Processing}, Brighton, UK, May 2019, pp. 416--420.

\bibitem{GoesslingWASPAA2019}
N.~G{\"o}{\ss}ling, W.~Middelberg, and S.~Doclo,
\newblock ``{RTF}-steered binaural {MVDR} beamforming incorporating multiple
  external microphones,''
\newblock in {\em Proc. IEEE Workshop on Applications of Signal Processing to
  Audio and Acoustics}, New Paltz, USA, Oct. 2019, pp. 368--372.

\end{thebibliography}

\end{document}